\documentclass[12pt]{article}
\usepackage{latexsym}
\usepackage{epsfig,amssymb,euscript,slashed}
\usepackage[linktocpage]{hyperref}
\usepackage{amsmath}
\usepackage{cite} 
\usepackage{array,calc,epsfig}
\usepackage{bbm}
\usepackage{fancybox}

\newcommand{\bra}{\langle}
\newcommand{\ket}{\rangle}

\numberwithin{equation}{section} 
\usepackage[]{graphicx}
\usepackage{color}

\begin{document}
\font\cmss=cmss10 \font\cmsss=cmss10 at 7pt

\begin{flushright}{
\scriptsize QMUL-PH-18-14}
\end{flushright}
\hfill
\vspace{18pt}
\begin{center}
{\Large 
\textbf{A note on the Virasoro blocks at order $1/c$ 
}}

\end{center}

\vspace{8pt}
\begin{center}
{\textsl{Alessandro Bombini$^{\,a,b}$, Stefano Giusto$^{\,a, b}$ and Rodolfo Russo$^{\,c}$}}

\vspace{1cm}

\textit{\small ${}^a$ Dipartimento di Fisica ed Astronomia ``Galileo Galilei",  Universit\`a di Padova,\\Via Marzolo 8, 35131 Padova, Italy} \\  \vspace{6pt}

\textit{\small ${}^b$ I.N.F.N. Sezione di Padova,
Via Marzolo 8, 35131 Padova, Italy}\\
\vspace{6pt}

\textit{\small ${}^c$ Centre for Research in String Theory, School of Physics and Astronomy\\
Queen Mary University of London,
Mile End Road, London, E1 4NS,
United Kingdom}\\
\vspace{6pt}

\end{center}

\vspace{12pt}

\begin{center}
\textbf{Abstract}
\end{center}

\vspace{4pt} {\small
  \noindent
  We derive an explicit expression for the $1/c$ contribution to the Virasoro blocks in 2D CFT in the limit of large $c$ with fixed values of the operators' dimensions. We follow the direct approach of orthonormalising, at order $1/c$, the space of the Virasoro descendants to obtain the blocks as a series expansion around $z=0$. For generic conformal weights this expansion can be summed in terms of hypergeometric functions and their first derivatives with respect to one parameter. For integer conformal weights we provide an equivalent expression written in terms of a finite sum of undifferentiated hypergeometric functions. These results make the monodromies of the blocks around $z=1$ manifest. }

\vspace{1cm}

\thispagestyle{empty}

\vfill
\vskip 5.mm
\hrule width 5.cm
\vskip 2.mm
{
\noindent  {\scriptsize e-mails:  {\tt alessandro.bombini@pd.infn.it, stefano.giusto@pd.infn.it, r.russo@qmul.ac.uk} }
}

\setcounter{footnote}{0}
\setcounter{page}{0}

\newpage


\section{Introduction}
Conformal symmetry is a powerful tool to constrain the correlators in two-dimensional CFT. It implies that four-point functions of primary fields on the sphere are given by (possibly infinite) sums of conformal blocks\footnote{We use the terms conformal and Virasoro block as synonymous.}, which are functions of the complex harmonic ratio: each block encodes the contributions of all the Virasoro descendants of a primary field. Given the central charge $c$ and the conformal dimensions $h_i$ and $h$ of the external and internal primary fields, conformal symmetry determines in principle the full functional dependence of the blocks on the harmonic ratio $z$. The infinite dimensionality of the 2D conformal algebra, which makes this powerful statement possible, also makes the task of computing conformal blocks particularly difficult. Multiple efforts have been devoted to this task since the seminal work of \cite{Belavin:1984vu}, but the general form of Virasoro blocks for generic values of $c$ remains unknown. Various perturbative expansions of the blocks can be generated via recursion relations \cite{Zamolodchikov:1985ie, Zamolodchikov:1988ij,Perlmutter:2015iya}. Combinatorial formulas for the coefficients in the $z$-expansion of the blocks have been found in \cite{Alba:2010qc}, based on the AGT correspondence \cite{Alday:2009aq} between 4d supersymmetric gauge theories and 2d CFT.

When the CFT admits a holographic dual, it is interesting to study the conformal blocks in the limit of large central charge. It is well known that in the $c\to \infty$ limit with fixed $h_i$ and $h$ -- a limit that we will denote as the LLLL limit -- the conformal blocks reduce to the global ones, associated with the projective subgroup of the local conformal group. This result can be extended in various directions: one can consider the so called HHLL limit, in which one keeps fixed the dimensions of the internal and of two (light) external operators and sends to infinity the dimensions $h_L$ of the two remaining (heavy) external operators, keeping the ratio $h_H/c$ fixed when $c\to \infty$. The leading contribution to the blocks in this regime has been derived in \cite{Fitzpatrick:2014vua,Fitzpatrick:2015zha}, and the subleading corrections have been studied in \cite{Beccaria:2015shq,Fitzpatrick:2015dlt,Chen:2016cms, Fitzpatrick:2016mjq}. One could also consider a classical limit in which all the dimensions $h_i$ and $h$ are rescaled together with $c$ \cite{Zamolodchikov:1995aa,Menotti:2014kra, Menotti:2016jut}. Conformal blocks have also been analysed from a bulk perspective exploiting their connection with geodesic Witten diagrams and Wilson lines \cite{Hijano:2015rla, Hijano:2015qja,  Hijano:2015zsa, Besken:2016ooo,Besken:2017fsj,Alkalaev:2015wia,Alkalaev:2015lca,Verlinde:1989ua,Fitzpatrick:2016mtp,Hikida:2017ehf,Hikida:2018dxe}. 

In this technical note we focus on the LLLL regime and derive exact expressions for the correction to the Virasoro blocks at order $1/c$. Our main motivation comes from holography, since this correction contributes to the connected part of correlators in the supergravity approximation. In Section~\ref{sec:pert} we compute the blocks by a direct method, summing over the Virasoro descendants that contribute at the desired order in the $c\to\infty$ limit. This produces the result \eqref{eq:blockh1} given as a series expansion in the ``direct channel'' $z\to 0$. In Section~\ref{sec:nonpert} we also make various attempts at summing the $z$-series to have access to the behaviour of the blocks also away from $z=0$. We first decompose the blocks into three contributions -- denoted as $f_a$, $f_b$, $f_c$ in \eqref{eq:fafbfc}  -- according to their dependence on the external dimensions $h_i$. Two first two terms, $f_a$ and $f_b$, were computed with the Wilson line approach in \cite{Fitzpatrick:2016mtp}, and we verify the agreement with our results. An explicit expressions for the last term was recently derived in~\cite{Hikida:2018dxe} by analytically continuing to a generic dimension $h$ the result for the $W_2$ minimal model obtained via the Wilson line approach developed in \cite{Hikida:2017ehf}. Here we provide two alternative exact expressions for $f_c$. The first \eqref{eq:blockgeneral} applies to generic real values of the internal dimension $h$, but involves derivatives of generalised hypergeometric functions with respect to one of their parameters. We also check that this result agrees with that of \cite{Hikida:2018dxe}, thus providing a further non-trivial confirmation of the Wilson-line approach at the quantum level. Then we find that for integer values of $h$ the block can be written as a finite sum of undifferentiated hypergeometric functions  (see \eqref{eq:blockinteger}). We also make some comments on the somewhat surprising singularity structure of the result: we find that the $1/c$ correction to the block has a leading singularity around $z=1$ that goes like $\log^2(1-z)$, as opposed to the $\log(1-z)$-singularity of the $c^0$ contribution. Since terms proportional to $\log^2(1-z)$ cannot arise in the expansion of the correlator in the crossed ($z\to 1$) channel, we speculate that these singularities have to cancel when an infinite series of conformal blocks is summed to produce a physical correlator.

\section{Perturbative Virasoro blocks at large $c$}
\label{sec:pert}

Let us consider a 2D CFT and focus on the correlator
\begin{equation}
  \label{eq:G1122}
  \langle O_1(z_1) O_1(z_2) O_2(z_3) O_2(z_4)\rangle =
  \frac{1}{z_{12}^{2h_1}\bar{z}_{12}^{2\bar{h}_1}} \frac{1}{z_{34}^{2h_2}\bar{z}_{34}^{2\bar{h}_2}}\; {\cal G}(z,\bar{z})\;,
\end{equation}
where $O_i$ are primary operators, $z_{ij}=z_i-z_j$, and $z$ is the following projective invariant cross ratio
\begin{equation}
  \label{eq:cross-ratio}
  z = \frac{z_{12} z_{34}}{ z_{13} z_{24}}\;. 
\end{equation}
We can use projective invariance to send $z_1\to \infty$, $z_2\to 1$ and $z_4\to 0$, so the cross ratio is identified with $z_3$. We denote with ${\cal P}_{h,\bar{h}}$ the projector on the subspace spanned by the Virasoro descendants of the primary state $|h,\bar{h}\rangle$. By inserting this projector in the correlator above, we isolate the contribution of a specific Virasoro block to the full correlator
\begin{equation}
  \label{eq:Vblhbarh}
 \langle O_1(\infty) O_1(1) {\cal P}_{h,\bar{h}} O_2(z) O_2(0)\rangle =
  C_{11h} C_{h22}\, z^{h-2 h_2} {\cal V}_h(z) \bar{z}^{\bar{h}- 2\bar{h}_2}\,\tilde{\cal V}_{\bar h}(\bar{z})\;,
\end{equation}
where $\langle O_1(\infty)\ldots \rangle = \lim_{z_1\to\infty} z_1^{2h_1} \bar{z}_1^{2\bar{h}_1}\langle O_1(z_1)\ldots \rangle $, $C_{iih}$ are the 3-point couplings between the exchanged operator $O_h$ and the external operators $O_i$ with $i=1,2$. The factor of $z^h$, $\bar{z}^{\bar{h}}$ are just a convention so as to normalise to $1$ the Virasoro blocks ${\cal V}_h(z)$, $\tilde{\cal V}_{\bar h}(\bar{z})$ in the $z,\bar{z} \to 0$ limit. 

The most naive approach to the derivation of ${\cal V}_h(z)$ is to try and construct the projector ${\cal P}_{h,\bar{h}}$ by using an orthonormal basis spanning the space of descendants of $|h\rangle$
\begin{equation}
  \label{eq:leftdesc}
  L_{-n}^{q_n}\ldots L_{-1}^{q_1}| h \rangle\;,
\end{equation}
where for notational simplicity we focused on the holomorphic sector. It is well known that this is a difficult task in general, but it is doable in perturbation theory in the large central charge limit. The reason is that the norm of the states in~\eqref{eq:leftdesc} is proportional to $c^{q_2+\ldots +q_n}$ and so the elements of the orthonormal basis are suppressed by a factor of $c$ for each Virasoro generator with mode lower than $-1$. Thus in the strict $c\to \infty$ limit, it is sufficient to focus on the descendants obtained by acting with $L_{-1}$, which implies that at leading order in $c$ the Virasoro blocks reduce to the the global conformal blocks. Here we are interested in the first subleading correction, so we need to consider the space spanned by descendants that have at most one $L_{-s}$ generator with $s\geq 2$. At level $q$ we have to deal with the following states
\begin{equation}
  \label{eq:subcspace}
  {\cal H}_q = \left\{ L_{-1}^q |h\ket \,,\,  L_{-2}  L_{-1}^{q-2}|h\ket\,,\, \ldots \,,\, L_{-s} L_{-1}^{q-s}  |h\ket\,,\, \ldots \,,\,  L_{-q} |h\ket \right\}\,.
\end{equation}
A convenient orthogonal basis is $| s \ket_q$, with $s=1,\ldots,q$, where
\begin{equation}
\label{eq_order1}
\begin{split}
| 1 \ket_q &= L_{-1}^q |h\ket \,, \\
| s \ket_q &= L_{-s} L_{-1}^{q-s}  |h\ket  - \sum_{j=1}^{s-1} \alpha_{(q,s)}^{(j)}(h) | j \ket_q \,~~\mbox{with}~~~s=2,\ldots,q\;. 
\end{split}
\end{equation}
Since ${}_q \langle s |s \rangle_q\sim c$ whenever $s>1$, all $\alpha_{(q,s)}^{(j)}$'s are of order $1/c$ except for $j=1$ where we have a coefficient $\alpha_{(q,s)}^{(1)}$ of order $1$. So at leading order the norm of $|s\rangle_q$ comes from its first term in~\eqref{eq_order1}
\begin{equation}
  \label{eq:normj}
  \begin{aligned}
    {}_q \langle 1 |1 \rangle_q & = q! (2h)_q \;, \\
    {}_q \langle s |s \rangle_q &= \langle h | L_1^{q-s} [L_s,L_{-s}] L_{-1}^{q-s}| h \rangle + O(c^0)\\
    &= \frac{c}{12} s (s^2-1) \langle h | L_1^{q-s} L_{-1}^{q-s}| h \rangle + O(c^0) \\
    &= \frac{c}{12} s (s^2-1)  (q-s)! (2h)_{q-s}  + O(c^0)\;,~~~s>1\;,
      \end{aligned}
\end{equation}
where we denote by $(h)_q$ the rising Pochhammer symbol
\begin{equation}
(h)_q\equiv \frac{\Gamma(h+q)}{\Gamma(h)}\,.
\end{equation}
Similarly we can calculate the leading part of the mixing coefficient $\alpha_{(q,s)}^{(1)} (h)$ that is the only one contributing at order $O(c^0)$
\begin{equation}
  \label{eq:alphaqs}
  \alpha^{(1)}_{(q,s)} (h) =  \frac{{}_q \bra 1 |L_{-s} L_{-1}^{q-s} |h\rangle}{{}_q \bra 1 |1 \ket_q}  = [h(s+1) +(q-s)]  \, \frac{(2h)_{q-s}}{(2h)_q}+O(c^{-1})\,.
\end{equation}
A simple way to compute the correlators appearing in \eqref{eq:normj} and \eqref{eq:alphaqs} is to use the standard conformal Ward identity (see for example Chapter 6 of \cite{DiFrancesco:1997nk}), which states that in a correlator containing primary operators at the points $z_i$ and a Virasoro descendant evaluated at $z$,  the operators $L_{-n}$ can be replaced by the differential operators $\mathcal{L}_{-n}$
\begin{equation}\label{eq:calLn}
L_{-n}\to \mathcal{L}_{-n} \equiv \sum_i \left[\frac{(n-1) h_i}{(z_i-z)^n}-\frac{\partial_{z_i}}{(z_i-z)^{n-1}}\right]\,.
\end{equation}
For $n=1$ it is convenient to write $\mathcal{L}_{-1}$ as $\partial_z$, which is equivalent to $-\sum_i \partial_{z_i}$ thanks to translation invariance. Acting with the appropriate ordered string of $\mathcal{L}_{-n}$'s on the two-point function $\langle O_h(z_1) O_h(z_2) \rangle$, and then taking $z_1\to\infty, z_2\to 0$, leads to the results in \eqref{eq:normj} and \eqref{eq:alphaqs}.

Then we can write the approximate projector on the Virasoro block of $O_h$ as
\begin{equation}
  \label{eq:Proj1/c}
  {\cal P}_{h} = \sum_{q=1}^\infty \frac{|1\rangle_q \; {}_q \bra 1 |}{{}_q \bra 1  |1 \ket_q} + \sum_{q=2}^\infty \sum_{s=2}^q \frac{|s\rangle_q \; {}_q \bra s |}{{}_q \bra s  |s \ket_q} + \ldots \;, 
\end{equation}
where, again, we are focusing just on the holomorphic part. The first term yields the leading (global) block
\begin{equation}
\label{eq:blockh0}
  {\cal V}_h^{(0)} (z) = {}_2 F_1(h,h;2h;z)\;.
\end{equation}
We now focus on the second term that captures the subleading $1/c$ correction ${\cal V}_h^{(1)}$
\begin{equation}
  \label{eq:Vhext}
  {\cal V}_h^{(1)} (z) =  z^{2 h_2 -h}\, \bar{z}^{2\bar{h}_2 - \bar{h}} \sum_{q=2}^\infty \sum_{s=2}^q \frac{\bra O_1 (\infty) O_1 (1) |s\ket_q \,  {}_q \bra s| O_2(z) O_2 (0) \ket }{C_{11h} C_{h22} \, {}_q\bra s|s\ket_q}  \,.
\end{equation}
Here we need to calculate the numerator and the denominator to the leading order, that is respectively at $O(c^0)$ and $O(c)$; the subleading correction in any of the two will contribute to ${\cal V}_h$ at order $O(c^{-2})$ and so can be neglected. To compute the three-point function
\begin{equation}\label{eq:3pointapp}
\bra  O_1 (\infty) O_1 (1)|s\ket_q = \bra  O_1 (\infty) O_1 (1) L_{-s} L_{-1}^{q-s} |h\ket - \alpha^{(1)}_{(q,s)}(h) \bra  O_1 (\infty) O_1 (1)L_{-1}^q |h\ket + O(c^{-1})\,,
\end{equation}
we can use again the operators \eqref{eq:calLn}, apply them on the three-point function 
\begin{equation}
\langle O_1(z_1) O_1(z_2) O_h(z_3) \rangle = \frac{C_{11h}}{z_{12}^{2h_1-h}z_{13}^h\, z_{23}^h}\,,
\end{equation}
and then take $z_1\to\infty$, $z_2\to 1$, $z_3\to 0$. This gives
\begin{equation}\label{eq:3point1}
\bra  O_1 (\infty) O_1 (1) L_{-s} L_{-1}^{q-s} |h\ket =C_{11h} \,\left[h_1 (s-1) + (h+q-s)\right] (h)_{q-s}
\end{equation}
and
\begin{equation}\label{eq:3point2}
\bra  O_1 (\infty) O_1 (1) L_{-1}^{q} |h\ket =C_{11h} \, (h)_{q}\,.
\end{equation}
One thus obtains, keeping only the $O(c^0)$ contribution in the large $c$ limit
\begin{equation}\label{eq:3point}
  \frac{\bra  O_1 (\infty) O_1 (1)|s\ket_q}{C_{11h}} \simeq \Big( \left[ h_1 (s-1) + (h+q-s) \right] (h)_{q-s} - (h)_q \,\alpha^{(1)}_{(q,s)} (h)\Big) \equiv B_{(q,s)} (h_1 ; h )\, .
\end{equation}
By using~\eqref{eq:alphaqs} for the mixing coefficient $ \alpha^{(1)}_{(q,s)}$ we have
\begin{equation}
B_{(q,s)} (h_i ; h ) =  \left\{ \left[ h_i (s-1) + (h+q-s) \right] (h)_{q-s} - [h(s+1) +(q-s)]  \, \frac{(2h)_{q-s}}{(2h)_q}  \, (h)_q \right\}\,.
\end{equation}
Similarly we find
\begin{equation}
\frac{ {}_q\bra s|  O_2 (z) O_L (0) \ket}{C_{h22}}  = z^{h-2 h_2 +q}\,\bar{z}^{\bar{h}-2 \bar{h}_2}\,  B_{(q,s)} (h_2 ; h ) \,.
\end{equation}
Thus we can write the $1/c$ correction ${\cal V}_h^{(1)} (z)$ to the full Virasoro block as follows
\begin{equation}
\label{eq:blockh1}
\begin{split}
{\cal V}_h^{(1)} (z) &= \frac{12}{c} \sum_{q=2}^\infty \left[ \sum_{s=2}^q  \frac{ B_{(q,s)} (h_1 ; h ) \, B_{(q,s)} (h_2 ; h ) }{s(s^2-1) (q-s)! \, (2h)_{q-s} } \right] z^{q}  \,.
\end{split}
\end{equation}
A closed form expression for all the coefficients in the $z$-expansion of ${\cal V}_h^{(1)} (z)$ was obtained recently in \cite{Hikida:2018dxe} (see eqs.~(4.40) and (4.42)), based on the approach developed in \cite{Hikida:2017ehf}. It is straightforward to check by using Mathematica that \eqref{eq:blockh1} agrees with the result of \cite{Hikida:2018dxe} as an expansion around $z=0$.

\section{Exact Virasoro blocks at large $c$}
\label{sec:nonpert}

The result~\eqref{eq:blockh1} allows one to easily derive the behaviour of the conformal block around $z=0$. It might be useful, in particular when applying the conformal bootstrap method, to have a non-perturbative control over ${\cal V}_h^{(1)} (z)$ away from $z=0$, and for this purpose one needs to sum the double series in~\eqref{eq:blockh1}. This is easily done for the identity block ($h=0$) and one obtains the result in eq.~(2.36) of \cite{Perlmutter:2015iya}: this is nothing but the global block of the stress tensor $T$, which is the only quasi-primary among the descendants of the identity that contributes at order $c^{-1}$. 

For general $h$ performing the summation in closed form is non-trivial. One can distinguish three terms in \eqref{eq:blockh1}, according to their dependence on the external conformal dimensions $h_1$ and $h_2$:
\begin{equation}
\label{eq:fafbfc}
{\cal V}_h^{(1)} (z) = \frac{12}{c} \left[ f_a(h;z) \,h_1 h_2 + f_b(h;z) (h_1+h_2) + f_c(h;z)\right]\,.
\end{equation}
Re-organising the sums as $\sum_{q=2}^\infty\sum_{s=2}^q\to \sum_{m=0}^\infty \sum_{s=2}^\infty$ with $m=q-s$, one finds
\begin{equation}
\label{eq:fa}
f_a(h;z)=\sum_{m=0}^\infty \sum_{s=2}^\infty \frac{z^{m+s}}{m!}\frac{s-1}{s(s+1)}\frac{(h)_m (h)_m}{(2h)_m}\,,
\end{equation}
\begin{subequations}
\label{eq:fb}
\begin{align}
f_b(h;z) &= f^{(1)}_b(h;z)-f^{(2)}_b(h;z)\,,\\
f^{(1)}_b(h;z)&=\sum_{m=0}^\infty \sum_{s=2}^\infty \frac{z^{m+s}}{m! \,s(s+1)}\frac{(h+m) (h)_m (h)_m}{(2h)_m}\,,\\
\label{eq:f2b}
f^{(2)}_b(h;z)&=\sum_{m=0}^\infty \sum_{s=2}^\infty \frac{z^{m+s}}{m!}\left[\frac{h+m}{s}-\frac{m}{s+1} \right] \frac{(h)_m (h)_{m+s}}{(2h)_{m+s}}\,,
\end{align}
\end{subequations}
\begin{subequations}
\label{eq:fc}
\begin{align}
f_c(h;z) &= f^{(1)}_c(h;z)-2 f^{(2)}_c(h;z)+f^{(3)}_c(h;z) \,,\\
f^{(1)}_c(h;z)&=\sum_{m=0}^\infty \sum_{s=2}^\infty \frac{z^{m+s}}{m! \,s(s^2-1)}\frac{[(h+m) (h)_m]^2}{(2h)_m}\,,\\
f^{(2)}_c(h;z)&=\sum_{m=0}^\infty \sum_{s=2}^\infty \frac{z^{m+s}}{m! \,s(s^2-1)}(h(s+1)+m)\,\frac{(h+m) (h)_m (h)_{m+s}}{(2h)_{m+s}}\,,\\
f^{(3)}_c(h;z)&=\sum_{m=0}^\infty \sum_{s=2}^\infty \frac{z^{m+s}}{m! \,s(s^2-1)}(h(s+1)+m)^2\,\frac{(2h)_m [(h)_{m+s}]^2}{[(2h)_{m+s}]^2}\,.
\end{align}
\end{subequations}
\bibliographystyle{utphys}      
Using the series representation of the generalised hypergeometric functions 
\begin{equation}
{}_p F_q(a_1,\ldots,a_p; b_1,\ldots,b_q;z) = \sum_{n=0}^\infty \frac{(a_1)_n \ldots (a_p)_n}{(b_1)_n\ldots (b_q)_n}\frac{z^n}{n!}\,,
\end{equation}
one sees that the $m$-series can be summed in terms of ${}_3F_2$ for the case of $f^{(3)}_c(h;z)$ and in terms of ${}_2F_1$ for all other cases. The remaining sum over $s$ can also be easily done for $f_a(h;z)$ and $f^{(1)}_b(h;z)$, where the $m$ and $s$-dependence factorise, and for $f^{(2)}_b(h;z)$, where one exploits a partial cancellation between the two terms in the square bracket in \eqref{eq:f2b}. One thus obtains:
\begin{equation} \label{eq:far}
f_a(h;z) =- \left(2+\frac{(2-z) \log(1-z)}{z}\right) \,{}_2F_1(h,h;2h;z)\,
\end{equation}
and
\begin{subequations}
\label{eq:fbab}
  \begin{align}
f^{(1)}_b(h;z) &= h\,\left(1+\frac{(1-z) \log(1-z)}{z}-\frac{z}{2}\right)\,{}_2F_1(h,h+1;2h;z)\,,\\
f^{(2)}_b(h;z) &= \frac{z^2}{2} \sum_{m=0}^\infty \frac{(h)_{m+1} (h)_{m+2}}{(2h)_{m+2}} \frac{z^m}{m!} = \frac{z^2}{4} \,\frac{h(h+1)}{2h+1}\,{}_2F_1(h+1,h+2;2h+2;z)\,.
\end{align}
\end{subequations}
The full $f_b(h;z)$ can be simplified to
\begin{equation}\label{eq:gb}
f_b(h;z) =  h \,{}_2F_1(h,h;2h;z) + h \frac{(1-z)\log(1-z)}{z}\,{}_2F_1(h,h+1;2h;z) \;.
\end{equation}
The above results for $f_a(h;z)$ and $f_b(h;z)$ agree with the ones derived with a different method in \cite{Fitzpatrick:2016mtp}. The same expressions can be derived by expanding at first order in $h_H/c$ the HHLL blocks at order $O(c^0)$ computed in \cite{Fitzpatrick:2015zha}.   

The genuinely new term is $f_c(h;z)$, and it is also the hardest to compute. In \cite{Fitzpatrick:2016mtp} the first few terms in the expansion around $z=0$ were given and it can be easily checked that these terms agree with \eqref{eq:blockh1}. Of the three contributions $f^{(i)}_c(h;z)$ with $i=1,2,3$, only $f^{(1)}_c(h;z)$ can be easily summed; $f^{(2)}_c(h;z)$ and $f^{(3)}_c(h;z)$ can be re-organised as a sum of series over $s$ containing hypergeometrics of the ${}_2F_1$ and ${}_3F_2$-type. All the series are of the form
\begin{equation}\label{eq:Fhatdef}
\begin{aligned}
&{}_{p+1}\hat F_q(a_1,\ldots,a_p,\alpha;b_1;\ldots,b_q;z)\equiv\\
&\quad\equiv \sum_{s=1}^\infty \frac{z^s}{s} \,\frac{\prod_p (a_p)_s}{\prod_q (b_q)_s} \, {}_{p+1} F_q (a_1+s,\ldots,a_p+s,\alpha;b_1+s,\ldots,b_q+s;z)\,,
\end{aligned}
\end{equation}
where $p=q=1$ for $f^{(2)}_c(h;z)$ and $p=q=2$ for $f^{(3)}_c(h;z)$; $a_i$, $b_i$ are functions of $h$. To sum these series we can use the identity \cite{Prudnikov:1990}
\begin{equation}
\begin{aligned}
\sum_{s=0}^\infty \frac{(\beta)_s\,z^s}{s!} \frac{\prod_p (a_p)_s}{\prod_q (b_q)_s} \,{}_{p+1}F_q(a_1+s,\ldots a_p+s, \alpha; b_1+s,\ldots b_q+s;z)=\\
\qquad\qquad\qquad\qquad={}_{p+1}F_q(a_1,\ldots a_p, \alpha+\beta; b_1,\ldots b_q;z)\,,
\end{aligned}
\end{equation}
and the fact that
\begin{equation}
\frac{d}{d\beta} \left (\beta)_s \right |_{\beta=0}=(s-1)!\,
\end{equation}
so that
\begin{equation}\label{eq:Fhatdefbis}
{}_{p+1}\hat F_q(a_1,\ldots,a_p,\alpha;b_1;\ldots,b_q;z)= \frac{d}{d\beta} \left.  {}_{p+1}F_q(a_1,\ldots a_p, \alpha+\beta; b_1,\ldots b_q;z) \right|_{\beta=0}\,.
\end{equation}
Collecting all the terms we arrive at a final general expression for $f_c(h;z)$:
\begin{align}
  \nonumber
f_c(h;z)&= -\frac{(h-1)^2}{2}\,{}_2F_1(h,h;2h;z)-\frac{h^2}{2}\frac{(1-z)^2\log(1-z)}{z}\,{}_2F_1(h+1,h+1;2h;z)\\ \nonumber
&-h^2 (z-2)\,{}_2\hat F_1(h+1,h+1;2h;z) -\frac{2 h (2h-1)}{z} \,{}_2\hat F_1(h,h+1;2h-1;z)\\ \nonumber
&+\frac{2 h (2h-1)}{z} \,{}_2\hat F_1(h-1,h+1;2h-2;z)\\
\label{eq:blockgeneral}
&+\frac{h^2}{2}(z-2)\,{}_3\hat F_2(h+1,h+1,2h;2h,2h;z)\\ \nonumber
&+\frac{2(2h-1)^2}{z} \,\Bigl({}_3\hat F_2(h,h,2h;2h-1,2h-1;z)\\ \nonumber
&+{}_3\hat F_2(h-1,h-1,2h;2h-2,2h-2;z)-2\,{}_3\hat F_2(h-1,h,2h;2h-2,2h-1;z)\Bigr) .
\end{align}
In Appendix \ref{sec:app} we show how this result can also be obtained from the integral formula of \cite{Fitzpatrick:2016mtp}. A formula that is similar in spirit to \eqref{eq:blockgeneral} follows from the approach of \cite{Hikida:2018dxe}:
\begin{equation}\label{eq:blockgeneralJ}
\begin{aligned}
f_c(h;z)=& -\frac{h^2}{2}\,{}_2F_1(h,h;2h;z)-\frac{h^2}{2}\frac{(1-z)^2\log(1-z)}{z}\,{}_2F_1(h+1,h+1;2h;z)\\
&+h(h-1) \left[{}_2 \hat F_1(h,h;2h;z)+{}_2\tilde F_1 (h,h;2h;z)\right]\,,
\end{aligned}
\end{equation}
with ${}_2 \hat F_1$ defined as in \eqref{eq:Fhatdefbis} and 
\begin{equation}
{}_2\tilde F_1 (a,b;c;z) \equiv \frac{d}{d\beta} \,{}_2F_1(a,b;c+\beta;z)|_{\beta=0}\,.
\end{equation}
We have checked that \eqref{eq:blockgeneral} and \eqref{eq:blockgeneralJ} agree for several values of $h$; this guarantees that the results of the Wilson-line approach are also valid non-perturbatively in $z$. 

As the behaviour of the generalised hypergeometric functions around $z=1$ is known for generic values of the parameters (see for example \cite{10.2307/2159793}), one could use either \eqref{eq:blockgeneral} or \eqref{eq:blockgeneralJ} to infer the singularity structure of general conformal blocks. In many applications, and especially in theories admitting holographic duals, it is useful to consider primaries with integer dimensions, and one might thus ask if a simpler form for $f_c$ could be obtained for integer $h$. The fact that $B_{(q,s)} (h_i=0 ; h=1)=0$ implies that $f_c(h=1,z)=0$. Inspecting the behaviour of $f_c$ for various integer values of $h\ge 2$, one finds that $f_c$ contains four type of terms: terms containing respectively $\log^2(1-z)$, $\mathrm{Li}_2(z)$, $\log(1-z)$ times a rational function of $z$, or terms containing only rational functions. The $\log^2$-terms only come from the last term in the first line of \eqref{eq:blockgeneral}. The $\mathrm{Li}_2$-terms come from the functions ${}_3\hat F_2$ and they can always be expressed as linear combinations of undifferentiated hypergeometric functions ${}_3F_2$, with constant coefficients: a closed form expression for these coefficients, for generic integer $h$, can be obtained by trial and error. Subtracting this linear combination from $f_c$, one can similarly generate the $\log$-terms of the remainder by a linear combination of ${}_2 F_1$: an educated guess is sufficient to determine all coefficients but one, which can be inferred by looking at the $z\to 0$ limit. Remarkably, one verifies that also the rational part of $f_c$ is reproduced. Finally we arrive at the following explicit expression for $f_c$, which we conjecture to be valid for any integer $h\ge 2$:
\begin{align}
f_c(h;z) &= -\frac{(h-1)^2}{2}\,{}_2F_1(h,h;2h;z)-\frac{h^2}{2}\frac{(1-z)^2\log(1-z)}{z}\,{}_2F_1(h+1,h+1;2h;z) \nonumber \\ \label{eq:blockinteger}
& \quad +  (-1)^h h (h-1)  \sum_{k=0}^{h-1} \frac{(h)_k  }{h+k} \, \frac{1}{k!} \, {}_3 F_2 ( h,h,h-k; 2h,2h  ; z) \\  \nonumber
& \quad + h  \sum_{k=1}^{h-1} (-1)^k \, \frac{{h+k-1 \choose h-1}}{ {h-2 \choose h-k-1} } \, {}_2 F_1 (h,h-k; 2h; z)    +  b(h)  \, {}_2 F_1(h,h;2h; z) \,,
\end{align}
with
\begin{equation}
b(h) = \frac{1}{2} - h + (-1)^{h-1} h\,(h-1) \sum _{k=0}^{h-1} \frac{(h)_k}{k! (h+k)} -h \sum _{k=1}^{h-1}(-1)^k \frac{ \binom{h+k-1}{h-1}}{\binom{h-2}{h-k-1}}  \  .
\end{equation}
With the help of the Mathematica package developed in \cite{Huber:2005yg}, we have checked that \eqref{eq:blockgeneral} and \eqref{eq:blockinteger} agree up to $h=10$ exactly in $z$. We have also checked that the expansion of \eqref{eq:blockinteger} around $z=0$ is in agreement with \eqref{eq:fc} for generic $h$ up to order $z^{20}$.

As mentioned at the beginning of this section, we can use these results to derive the monodromies around $z=1$ of the Virasoro block at order $1/c$ from the knowledge of the non-analytic behaviour of the hypergeometric functions. This is particularly straightforward in the case of integer conformal weights where we can use~\eqref{eq:blockinteger} (together with~\eqref{eq:far} and~\eqref{eq:gb} which are valid in general). A new feature of the $1/c$ corrected Virasoro block is the presence of terms proportional to $\log^2 (1-z)$, which are absent in the global blocks. Using known results for the non-analytic behaviour of ${}_2F_1$ \cite{Lebedev:72407}, one finds that the $\log^2(1-z)$ terms of $f_a$, $f_b$, $f_c$ are 
\begin{equation}\label{eq:abclog2}
\begin{aligned}
&f_a(h;z)\simeq \frac{\Gamma(2h)}{\Gamma^2(h)}\,\frac{2-z}{z}\,{}_2F_1(h,h;1;1-z)\log^2(1-z)\,,\\
&f_b(h;z)\simeq h (h-1) \frac{\Gamma(2h)}{\Gamma^2(h)}\,\frac{1-z}{z}
\,{}_2F_1(h,h+1;2;1-z)
\log^2(1-z) \,,\\
&f_c(h;z)\simeq \frac{h^2(h-1)^2}{4} \frac{\Gamma(2h)}{\Gamma^2(h)}\,\frac{(1-z)^2}{z}\,{}_2F_1(h+1,h+1;3;1-z)\log^2(1-z)\,.
\end{aligned}
\end{equation}
The large $c$ expansion of a CFT correlator generically contains terms with logarithms, which are related to the $1/c$ corrections to the conformal dimensions of multiparticle operators. However there is no room, at order $1/c$, for terms with the logarithm squared such as the ones appearing in~\eqref{eq:abclog2}. This probably means that in a large $c$ 2D CFT the correlators always involve an infinite number of Virasoro primaries so as to avoid the appearance of the $\log^2$ that are present in each block. If possible, it would of course be very interesting to generalise what has been done by using the global blocks (see for instance~\cite{Heemskerk:2009pn,Alday:2017gde}) to the case of the Virasoro blocks and exploit the full Virasoro algebra in the holographic reconstruction of AdS$_3$ physics.

\section*{Acknowledgements}

We would like to thank L. F. Alday and A. Bissi for discussions and correspondence. This work was supported in part by the Science and Technology Facilities Council (STFC) Consolidated Grant ST/L000415/1 {\it String theory, gauge theory \& duality}.

\appendix
\section{Relation with the Wilson line approach}
\label{sec:app}

In~\cite{Fitzpatrick:2016mtp} the authors used the Wilson line approach~\cite{Verlinde:1989ua} to provide a compact integral expression for the large $c$ Virasoro block, see eqs.~(3.19) and~(3.20) of the paper mentioned above\footnote{Eq.~(3.20) of ~\cite{Fitzpatrick:2016mtp} should have an extra factor of $z_{21}^{-1} z_{43}^{-1}$ and the overall factor of~(3.24) should read $1/24$ instead of $1/2$.}. It is straightforward to show that from this formulation one can find the same type of series that we encountered in the main text and that can be summed by using~\eqref{eq:Fhatdef}.

The strategy is as follows: eq.~(3.20) of~\cite{Fitzpatrick:2016mtp} involves four integrals (over the two punctures $z_5$ and $z_6$ of the exchanged state $O_p$  and over the variables $w_1$, $w_2$ included in the definition of ${\cal T}$); one can use the standard OPE expansions of $O_p(z_5) O_p(z_6)\sim z_{56}^{-2h_p}$ and the stress tensor appearing in  $\langle {\cal T}{\cal T} \rangle$ to write the integrand in~(3.20) explicitly; finally one can fix a gauge for the position of the external states as done in Section~\ref{sec:pert}, perform the integrations over $w_1$, $w_2$, which are straightforward, and rewrite the remaining two integrals in terms of the variables $y_5=z_5/z$ and $y_6=1/z_6$ which both run in the interval $[0,1]$.

The integrals for the terms that depend on the conformal weights $h_1$, $h_2$ of the external states have been already performed in~\cite{Fitzpatrick:2016mtp} and correspond to the contributions in~\eqref{eq:fa} and~\eqref{eq:fb}. Thus we can focus on the part of the integrand independent of $h_1$ and $h_2$
\begin{equation}
  \label{eq:y5y6int}
  \begin{aligned}
    f_c(h;z) = &  \frac{(h-1)}{2 z} \frac{\Gamma^2(2h)}{\Gamma^4(h)} \int\limits_0^1 \! dy_5 \int\limits_0^1 \! dy_6\, (1-y_5)^{h-2} y_5^{h-2} (1-y_6)^{h-2} y_6^{h-2} (1-y_5 y_6)^{-2h} \\
    \times &~ \Big\{- y_5 y_6 \left[(1-y_5) (1-y_6) z + y_5y_6 (1-z)^2 \log(1-z)\right] + \\
    &~~~~ y_5^2\,\left[y_6^2 (z-2) z+2 y_6-1\right] \log (1-y_6 z) +  \\
    &~~~~ y_6^2\,\left[y_5^2 (z-2) z+2 y_5-1\right] \log (1-y_5 z)+ \\
    &~~~~ \left[-y_5^2 y_6^2 (z-2) z - 4 y_5 y_6 + 2 (y_5+y_6)-1\right] \log (1- y_5 y_6 z) \Big\}\;.
  \end{aligned}
\end{equation}
We notice that it is divided in three different types of terms: contributions proportional to $\log(1-y_5 y_6 z)$, the ones proportional to either $\log(1-y_5 z)$ or $\log(1-y_6 z)$ and those without any such logarithms. In the latter case the integration over $y_5$ and $y_6$ can be performed straightforwardly and, by using the integral definition of the Gauss hypergeometric function, one obtains the first line of~\eqref{eq:blockgeneral}. The terms proportional to $\log(1-y_5 z)$ and $\log(1-y_6 z)$ yield the same result: it is convenient to start integrating the variable that is not present in the logarithm, then expand the logarithm in series for small $y_i$ and integrate each term. In this way one obtains exactly a series of the type of~\eqref{eq:Fhatdef}, which generates the terms proportional to ${}_2\hat F_1$ in~\eqref{eq:blockgeneral}. Finally one can treat in a similar fashion also the terms in the integrand proportional to $\log(1-y_5 y_6 z)$. After expanding the logarithm in series, the first integral (over $y_5$ for instance) yields terms with a Gauss hypergeometric function and then the second one (over $y_6$) is a Euler type of integral that yields a generalised hypergeometric ${}_3 F_2$. All coefficients conspire to produce a series of the type of~\eqref{eq:Fhatdef}, so the final result can be written in terms of ${}_3\hat F_2$. By combining all the contributions one obtains~\eqref{eq:blockgeneral}.

\providecommand{\href}[2]{#2}\begingroup\raggedright\endgroup

\end{document}